\documentclass[letterpaper, 10 pt, conference]{ieeeconf}  

\IEEEoverridecommandlockouts                              

\usepackage[linesnumbered,ruled,vlined]{algorithm2e}
\usepackage{amssymb}
\usepackage{mwe} 

\title{
\begin{center}
\vspace{-1.5cm} \footnotesize Accepted for presentation at the 43rd Annual International Conference of the IEEE Engineering in Medicine \& Biology Society (EMBC'2021)
\vspace{1.0cm}
\end{center}

\LARGE \bf A Similarity Measure of Histopathology Images by Deep Embeddings}
\author{Mehdi Afshari, H.R. Tizhoosh
\thanks{M. Afshari and H.R. Tizhoosh are members
of the Laboratory for Knowledge Inference in Medical Image Analysis
(Kimia Lab), University of Waterloo, Waterloo, ON, Canada, (e-mail:
\{m4afshar, tizhoosh\}@uwaterloo.ca).}
}
\begin{document}
%

\maketitle
\begin{abstract}
Histopathology digital scans are large-size images that contain valuable information at the pixel level. Content-based comparison of these images is a challenging task. This study proposes a content-based similarity measure for high-resolution gigapixel histopathology images. The proposed similarity measure is an expansion of cosine vector similarity to a matrix. Each image is divided into same-size patches with a meaningful amount of information (i.e., contained enough tissue). The similarity is measured by the extraction of patch-level deep embeddings of the last pooling layer of a pre-trained deep model at four different magnification levels, namely, 1x, 2.5x, 5x, and 10x magnifications.  In addition,  for faster measurement, embedding reduction is investigated.  Finally, to assess the proposed method, an image search method is implemented. Results show that the similarity measure represents the slide labels with a maximum accuracy of 93.18\% for top-5 search at 5x magnification.
\end{abstract}
\section{Introduction}
\label{sec:intro}

Histopathology images are gigapixel files containing a considerable amount of information on details of tissue structure.  Measurement of the similarity of tissue morphology can be advantageous to facilitate research, educational and diagnostic applications. In digital pathology, these gigapixel images, called whole slide images (WSIs), are commonly split into patches for further assessment \cite{babaie2017classification,kalra2020yottixel}. This division is helpful, among others, to address the processor-memory bottleneck issues. Patches are extracted by multiple strategies where the size of each patch, degree of their overlap, and the respective number of patches may vary for different purposes. One of the popular methods is to extract patches with no overlapping \cite{zhu2018multiple}. The WSIs have a pyramid structure that contains multiple magnifications \cite{goode2013openslide}. By extraction of patches, each WSI is described by numerous small images at the chosen magnification. Selection of the magnification is still a challenging task. Digital pathology researchers tackle the problems at different magnifications \cite{coudray2018classification}.

During the past decade, numerous deep learning applications for digital pathology have been studied \cite{janowczyk2016deep}. Convolutional Neural Networks (CNNs) are the main structure of deep learning to learn and extract deep embeddings, i.e., salient features, of the images. Pre-trained CNNs on \emph{ImageNet}, a general-purpose and widely used image dataset \cite{deng2009imagenet}, are used to extract image features. ResNet-50 is one of the successfully trained networks to extract descriptive features \cite{he2016deep}. The pre-trained networks are mostly used to provide descriptive features of the patches  \cite{gildenblat2019self}. These features can serve as a concise image representation. In the area of medical imaging, numerous studies have used pre-trained networks \cite{deniz2018transfer}. Networks pre-trained using ImageNet are commonly used in a wide range of domains due to their performance and availability. Measurement of ``image similarity'' has been subject of many studies \cite{wang2020image} \cite{coltuc2018use}. These measures are often used to quantify the similarity mostly via the distance of two images \cite{wang2016semantic}. The proper measures are generally constructed in a manner to support multiple objectives. Firstly, they should be larger than zero. Secondly, if the two similar images are duplicates, they should have zero distance to show full similarity. Finally, the distance of the two different images should be the same regardless of the comparison order \cite{Mitchell2010}. In this study, a new measurement scheme is proposed which is capable of similarity comparison of WSIs when patches are extracted.

The paper is organized as follows: In section \ref{sec:related_works},  related works are briefly reviewed. In section \ref{sec:method}, the proposed method is described containing three major sub-parts of extraction and reduction of features which is aimed to enhance the computaion speed, construction of similarity matrix for two WSIs, and finally the image search method. In  section \ref{sec:exp}, the conducted experiments including the dataset and implementations are described. Results of similarity measurements for the purpose of image search are analyzed subsequently. Section \ref{sec:conc} conclude the paper.

\section{Related Works}
\label{sec:related_works}

The measurement of image similarity has been studied in other domains for rather small images. One of the first developments in this area is presented by IBM in a content-based image retrieval system called QBIC (Query By Image Content) \cite{niblack1993qbic}. Their method is based on texture, shape, and color to retrieve similar images. However, QBIC was capable of comparing only small size images. As well, there was a considerable trade-off between local and global information. In \cite{alsmadi2017efficient}, the authors studied the efficiency of a similarity measure based on a genetic algorithm. Their method benefits from color and shape features. Though their method is quite robust, the variety of WSIs may confuse the algorithms due to variations in shape and staining.  

In medical imaging, Qayyum et al. presented a study of retrieval using deep CNNs \cite{qayyum2017medical}. Their main focus is on finding a solution to minimize the gap between semantic information in an image and numerical features. Their study was performed on CT images which are usually considerably smaller than digital pathology images. Although they achieved high accuracy in retrieval, their method is limited to use for small size images. In  histopathology, only a few studies have embarked upon image similarity measurements for gigapixel images. Among those few, Gildenblat presented a self-supervised method of image similarity learning that employs ResNet-50 but it does not offer any finite valued measure for images as a comparison \cite{gildenblat2019self}. 

\section{Method}
\label{sec:method}

\begin{figure*}[htb]
    \centering
    \centerline{\includegraphics[width=1\linewidth]{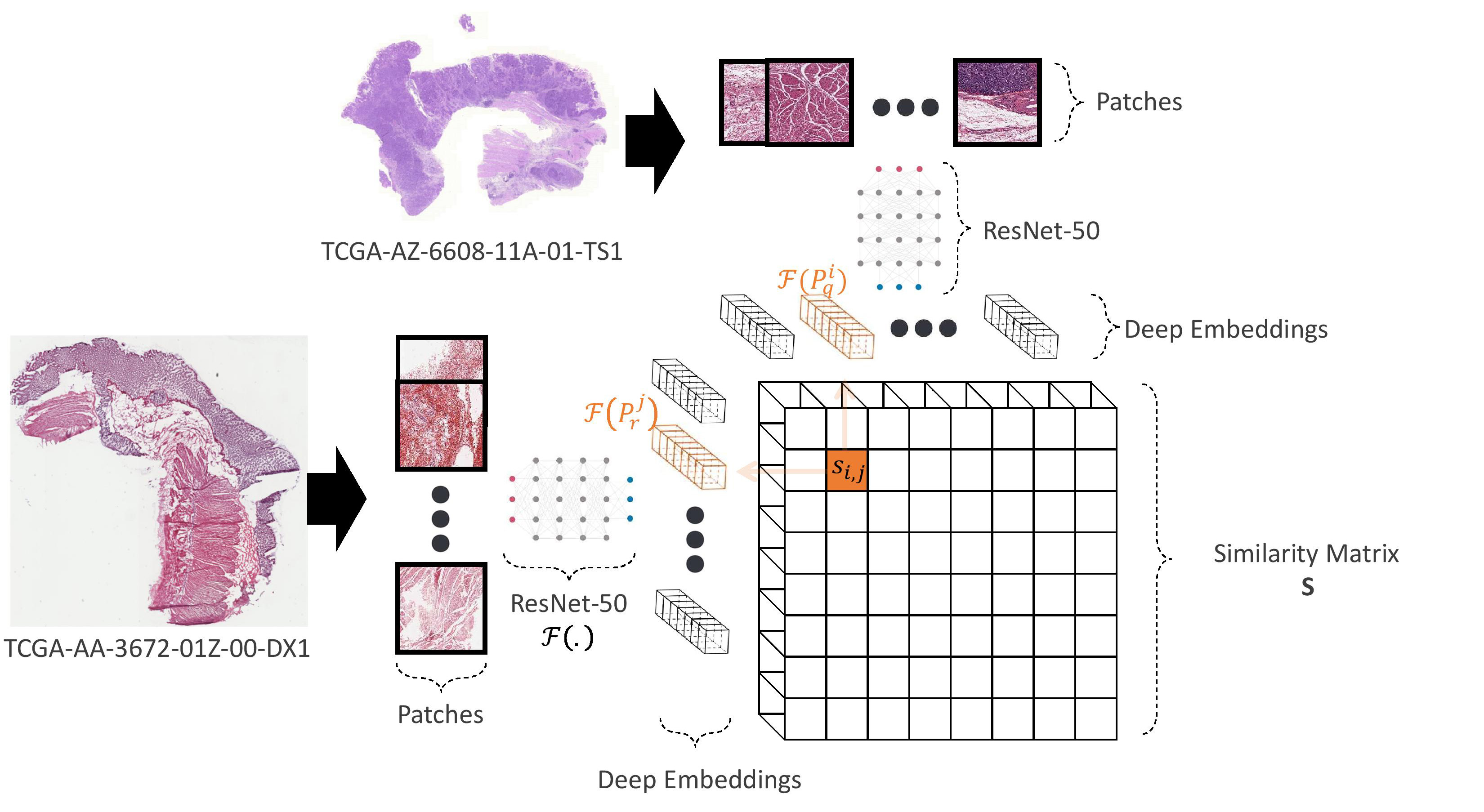}}
    \caption{Overall structure of  similarity measurement. The patches are fed to a deep network. Deep embeddings are used to construct the similarity matrix.}
    \label{overall}
\end{figure*}

The overall procedure of the proposed method is depicted in Fig. \ref{overall}. A query WSI is a considerably large image as an input. The size of these images is much larger than the images of \emph{natural} images in ImageNet dataset commonly used in machine-learning research. The embeddings of these images should be extracted by using a pre-trained network (e.g., ResNet-50); therefore, each WSI is first split into the same size sub-images called patches. Subsequently, background patches are detected and dropped by a method described later. Then, embeddings of the last pooling layer of the pre-trained ResNet-50 are extracted. Next, the similarity matrix of the query image and a reference image is computed. Finally, most similar images are selected based on the introduced similarity measure computed based on the similarity matrix.

\begin{figure}[htb]
    \centering
    \centerline{\includegraphics[width=0.9\linewidth]{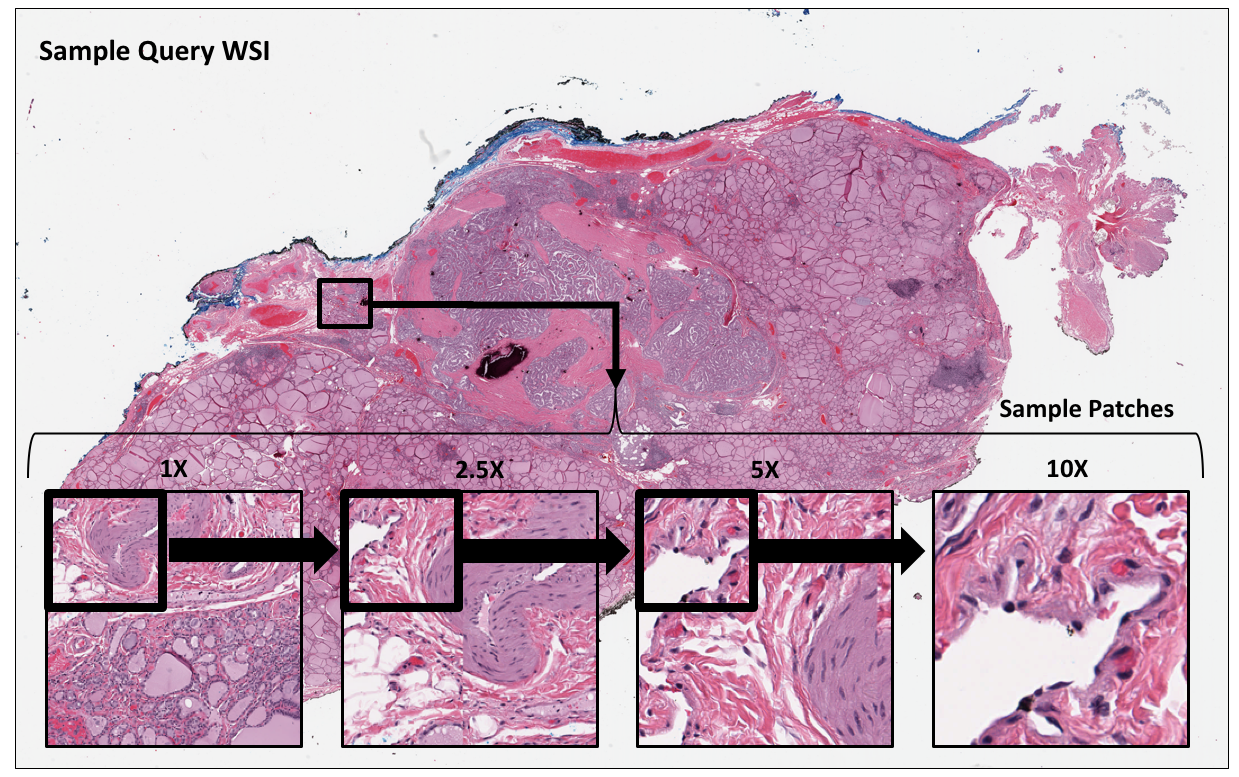}}
    \caption{A WSI query and sample patches  at different magnifications.}
    \label{sample_patch}
\end{figure}

\textbf{Feature Extraction and Reduction --} 
Patches are used to extract deep embeddings by the last pooling layer of the pre-trained ResNet-50; therefore, for each patch, $\mathbf{P} \in \mathbb{R}^{s_p} \times \mathbb{R}^{s_p} \times \mathbb{R}^{3}$, a vector of features, $\mathbf{d} = \mathcal{F}(\mathbf{P}) \in \mathbb{R}^{s_f}$, presenting the embeddings of the patch is extracted. Where $\mathcal{F}(\cdot)$ computes the output of last polling layer of the pre-trained ResNet-50.

As a faster approach, the dimensionality reduction of the extracted features has been exercised. This procedure is followed for each magnification level, however, for the sake of simplicity, calculations are described regardless of magnification for a subset of $n$ patches. Consequently, the mean deep embedding, $\overline{ \mathbf{d} } = \frac{1}{n}\sum_i \mathbf{d}^i$, is computed where $\mathbf{d}^i$ is the deep embedding of the $i$-th patch. Next, the standard deviation of the deep embedding elements, $\mathbf{\sigma} =\sqrt{\frac{1}{n} \sum_{i} (\mathbf{d}_i-\overline{\mathbf{d}})}$, is calculated. Finally, the vector of coefficient of variations, $\mathbf{v}$, is computed where its elements are $v_j = {\sigma_j} / {\overline{d}_j}$. The vector, $\mathbf{v}$, is eventually sorted from high to low. The top $n_f$ feature values are selected and the rest of the elements of the vector are eleminated. It worth noting that the elmination is applied on the respective elements of all of the deep embeddings.

\textbf{Proposed Similarity Measurement --} 
In this section, the similarity measure that consists of a developed similarity matrix is explained. The similarity measure is the mean of two values calculated by averaging of the maximum of the similarity matrix and it's transposed matrix. On the other hand, the similarity matrix elements are the cosine similarity of patches of the patches. The calculation of the measure is as follows. For the sake of simplicity of description, the calculation steps are described regardless of magnification level, while in implementation, same procedure for each magnification level has been practiced. Without loss of generality, we assume that two images as query, $\mathbf{I}_q$, and reference, $\mathbf{I}_r$, are compared. The number of the patches of the query and reference image are assumed as $n_q$ and $n_{r}$, respectively. Besides, the $i$-th patch of the query image is denoted as $\mathbf{P}_q^i$ and the $j$-th patch of the reference image is denoted by $\mathbf{P}_r^j$. The similarity matrix, $\mathbf{S} \in \mathbb{R}^{n_q} \times \mathbb{R}^{n_r}$, is computed as 

\begin{equation}
    \label{eqn::sim_mat}
    s_{i,j} = \frac{\mathcal{F}(\mathbf{P}_q^i) \cdot \mathcal{F}(\mathbf{P}_r^j)} {||\mathcal{F}(\mathbf{P}_q^i)||_2\times||\mathcal{F}(\mathbf{P}_r^j)||_2},
\end{equation}

where $||.||_2$ denotes the $L_2$ norm.

Finally, the similarity of the query image and the reference image, $sim(\mathbf{I}_q, \mathbf{I}_r)$, is measured as 

\begin{equation}
    \label{eqn::sim_fcn}
    sim(\mathbf{I}_q, \mathbf{I}_r)\! =\! \frac{1}{2}\! \times\! \Bigg [
    \frac{1}{n_r} \sum_{j} \max_{i} s_{i,j}  
    + \frac{1}{n_q} \sum_{i} \max_{j} s_{i,j}  \Bigg ].
\end{equation}

\textbf{Properties --} Based on the provided equation in 1, it is evident that the elements of the similarity matrix are values in $s_{i,j} \in [0,1]$. Consequently, the similarity measure is also in the range positive real-valued numbers smaller than one. If two similar images are compared, same patches are compared; therefore, the similarity matrix will contain a diagonal of ones. Based on \ref{eqn::sim_fcn}, it ensures that the similarity measure is also equal to one. Besides, computation of similarity measure by \ref{eqn::sim_fcn} results in the same measure value for query to reference image and wise-verse.

\begin{table*}[t]
\caption{Results of the Experiments}
\label{res}
\centering
\begin{tabular}{c|c|c|c|c|c|c|c|c}
                                & 1X             & 2.5X & 5X & 10X & 1X-128 & 2.5X-128 & 5X-128 & 10X-128 \\
\hline \hline
CPU Runtime                     &  \textless 1 min. & 3 mins    & 10 mins & 35 hours & \textless 1 min. & 2 mins                 & 8 mins               & 30 hours              \\
\hline
GPU Runtime                     &  \textless 1 min. & 2 mins    & 5 mins  & 30 mins  & \textless 1 min. & 1 min                  & 4mins                & 25 mins               \\
\hline
Top-3 Accuracy  & 82.95\%             & 77.27\%   & 82.95\% & 73.86\%  & 80.68\%              & 68.18\%                & 82.95\%              & 73.86\%               \\
\hline
Top-5 Accuracy                  & 90.91\%             & 89.77\%   & 93.18\% & 86.36\%  & 92.05\%              & 82.95\%                & 90.91\%              & 86.36\%             
\end{tabular}
\end{table*}

\textbf{Image Search --} 
A sample query image is compared with other images to find the most similar images based on the introduced measure to search. The leave-one-out strategy has been taken into practice for image search. The query image is first patched, and then informative patches are extracted. The features of the patches by the use of pre-trained ResNet-50 are extracted, and then the selected features are processed by the similarity measurement method. The final values of the similarities are sorted from high to low, and the top-3 and top-5 most similar images which produced the highest similarity measure are chosen as the search result. A search result is considered correct when at least one image with the same label of the query image is selected. Finally, the accuracy shows the rate of correct search result occurrence.

\section{Experiments}
\label{sec:exp}

In this study, we use a pre-trained ResNet-50 network. The output of the last polling layer of the network is the deep embedding of the corresponding patch. Consequently, the patch size, $s_p$, is $224\times224$ pixels in 3-channels of the RGB color-space to match the training image size of the original network. Besides, a deep embedding, $\mathbf{d}$, contains $s_f = 2048$ elements while the reduced version is a vector of, $n_f=128$, length real-valued elements. Finally, the images are stored in 8-bit color graphics which results in $c_{min}=0$ and $c_{max}=255$. Description of the dataset and implementation and results are presented in next parts. 

\textbf{Dataset --} To test the proposed measure, a popular histopathology dataset, namely, The Cancer Genome Atlas (TCGA) (available at https://www.cancer.gov/tcga) is used \cite{weinstein2013cancer}. The TCGA dataset carries digital slide scans of 25 different anatomical sites which includes 32 different cancer sub-types. In this study, a 20\% randomly selected images of the gastrointestinal tract anatomic site of the TCGA dataset are evaluated. It includes 88 slides with the distribution shown in Table \ref{table_dr}. The procedure that results in extraction of tissue patches is described in \ref{method:patch_extract} and \ref{method:outliers}.
 
\subsubsection{Patch Extraction}
\label{method:patch_extract}

 The patching is done at 4 different magnifications $m$, namely at 1x, 2.5x, 5x, and 10x magnifications. Fig. \ref{sample_patch} shows a sample query image, where one patch at each of the four magnification levels of the image are shown. Consequently, for a desired patch size of $s_p$, a sample WSI is divided into a total of $n^m = \lfloor \frac{h}{s_p} \rfloor \times \lfloor \frac{w}{s_p} \rfloor$ patches at the $m$ magnification levels where $h$, and $w$ are the pixel-wise height and width of WSI, respectively.

\subsubsection{Background patches removal}
\label{method:outliers}

 When WSIs are split into small patches a considerable number of patches may cover only the background pixels. Background patches are identified via binned histograms. Three histogram bins are $[c_{min},a]$, $(a,b]$, and $(b,c_{max}]$ where $a$ and $b$ are the free parameters that have been chosen by minimization of appearance of a random subset of manually labeled patches, respectively. The constants $c_{min}$ and $c_{max}$ are the minimum and maximum pixel color-space values, respectively. The outlier patches are then removed when the ratio of the third to first histogram bin is smaller than a constant. The remaining patches are used in next steps. The number of patches at each magnification is updated accordingly.

\begin{table}[t]
    \vskip0.2cm
    \caption{Distribution of the randomly selected WSIs}
    \label{table_dr}
    \begin{center}
        \begin{tabular}{c||c|c}
        TCGA Code & Primary Diagnosis     & Slides \\
        \hline \hline
        COAD & Colon Adenocarcinoma & 14 \\
        \hline
        ESCA & Esophageal Carcinoma  & 32 \\
        \hline
        READ & Rectum Adenocarcinoma & 30 \\
        \hline
        STAD & Stomach Adenocarcinoma & 12 \\
        \end{tabular}
    \end{center}
\end{table}

\textbf{Implementation --} The method has been implemented and evaluated on an Nvidia-2080 Super GPU and 10 cores Intel 9900X CPU. The GPU computations represent the computation of deep embeddings while the rest of similarity measurement method requires CPU computation.

\begin{table}[t]
    \caption{Distribution of extracted patches}
    \label{dep}
    \begin{center}
        \begin{tabular}{c||c|c}
        Magnification & tissue patches     & background \\
        \hline \hline
        1X & 4668 & 1243 \\
        \hline
        2.5X & 22989  & 9863 \\
        \hline
        5X & 84596 & 25672 \\
        \hline
        10X & 319563 & 75689 \\
        \end{tabular}
    \end{center}
\end{table}

\textbf{Results --} A total number of 431K patches for all magnification levels are extracted for the slides. The detailed total count of standard and non-informative patches at each magnification level are explained in the Table  \ref{dep}. 
The results of the accuracy of the measure in image search are presented as the Table \ref{res} where 1X , 2.5, 5X , and 10X refer to the search by measurement of the similarity of the WSIs in 1X, 2.5X, 5X, 10X magnifications with all 2048 extracted feature values, respectively. The rest of the table as 1X-128, 2.5-128, 5X-128, and 10X-128 shows the results of the search by measurement of the similarity of the WSIs in 1X, 2.5X, 5X, 10X magnifications with reduced 128 feature values, respectively. To compare the performance of the method at different magnification levels, four measures of GPU/CPU runtimes, top-3 and top-5 accuracy values are presented.

\section{Discussion and Conclusions}
\label{sec:conc}

In this study, a measure of similarity for high-resolution images is presented. In this method, the similarity of two slides is computed in percentages. The measure uses pre-trained ResNet-50, which helps in content-based similarity and fast performance in the GPU/CPU runtime for processing. An image search chain has been implemented to assess the similarity values for categorizing four subsets of the TCGA dataset. The results showed that the method could measure the similarity of images where it reached the top-5 accuracy of 93.18\% for 5X magnification. These results show the correlation of measure and the labels. Overall, the magnification of 5X performed better than other magnifications. The method results also showed that the reduction of features might adversely affect the method's performance from an accuracy perspective. However, it accelerates the computation. The other benefit of the feature reduction could be a broader range of similarity measurement values. The proposed approach maps the similarity of images to a $[0-100\%]$ range enabling thresholding the search results according to the preferences of clinical experts. 

\vspace{0.15in}
\textbf{Ethics Statement --} No human subjects or animals were used for this research. Only secondary data from public TCGA dataset have been used. 




\bibliographystyle{Main}
\bibliography{Main}

\end{document}